# BEYOND VISUAL P300 BASED BRAIN-COMPUTER INTERFACING PARADIGMS

**T. M. Rutkowski**

*Life Science Center of Tara, University of Tsukuba, Ibaraki, Japan, tomek@tara.tsukuba.ac.jp*

## ABSTRACT

The paper reviews and summarizes recent developments in spatial auditory and tactile brain-computer interfacing neurotechology applications. It serves as the latest developments summary in “non-visual” brain-computer interfacing solutions presented in a tutorial delivered by the author at the IICST 2013 workshop. The novel concepts of unimodal auditory or tactile, as well as a bimodal combined paradigms are described and supported with recent research results from our BCI-lab research group at Life Science Center, University of Tsukuba, Japan. The newly developed experimental paradigms fit perfectly to needs of paralyzed or hearing impaired users, in case of tactile stimulus, as well as for able users who cannot utilize vision in computer or machine interaction (driving or operation of machinery required not disturbed eyesight). We present and review the EEG event related potential responses useful for brain computer interfacing applications beyond state-of-the-art visual paradigms. In conclusion the recent results are discussed and suggestions for further applications are drawn.

## 1. INTRODUCTION

Brain computer interfaces (BCI) are the emerging human-machine interaction technologies that require only user’s intentional brain activity modulation to generate commands (Walpow and Walpow, 2012). Adaptive machine learning applications (Krusienski et al., 2006) are at the center of such interactive neurotechnolgy applications. Carefully designed BCI interactive paradigms are expected to allow for an online non-invasive monitoring and further classification of human brain wave patterns to be later translated to the computer or machine control commands. By definition the BCI applications do not require any human muscle movement activities usually necessary in the classical human-computer-interfacing (HCI). For BCI the monitored and adaptively classified brain wave patterns (usually preceding any behavior actions) are only necessary. This allows for creating a novel and intuitive interactive interfacing technology for paralyzed or totally locked-in (TLS) patients who cannot execute any bodily movements. The healthy users could also benefit from the BCI technology in smart environments or vehicular robot control, computer gaming, or bodily training/rehabilitation technologies. The vast majority of contemporary BCI applications relay on movement imagery or visual stimuli paradigms (Walpow and Walpow, 2012), which are known to require from the users heavy attention resources. The auditory and tactile sensory modalities are recently a focus of very active research in neurotechnology (Cai et al., 2012; Cai et al., 2013; Chang et al., 2013; Mori et al., 2013a; Mori et al., 2013b; Rutkowski et al., 2013).

Two BCI modality paradigms are described in this paper which result with very encouraging classification accuracies bringing a possibility a real world application very soon. Non-invasive BCI neurotechnology applications are usually based on the monitoring of brain electrical waves by means of the electroencephalogram (EEG) (Walpow and Walpow, 2012). Owing to its non-invasive nature and application simplicity, the EEG based BCI are the best candidates to be at the core of future “intelligent” interfaces, prosthetic and neuro-rehabilitation devices. A concept of utilizing brain auditory or somatosensory (tactile) modalities creates a very interesting possibility to target “the less crucial” sensory domains, which are not as demanding as vision in operation of machinery or vision based computer applications (see Figure 1 with an example of a subject trying to drive a wheelchair while pay full visual attention at a computer display with the visual stimuli). Auditory and tactile BCI are thus potentially less mentally demanding receiving recently more attention in neurotechnology applications (Brouwer and Van Erp, 2010; Mori et al., 2012; Cai et al., 2012; Ortner et al., 2012; van der Waal et al., 2012; Cai et al., 2013; Chang et al., 2013; Mori et al., 2013a; Mori et al., 2013b; Rutkowski et al., 2013).

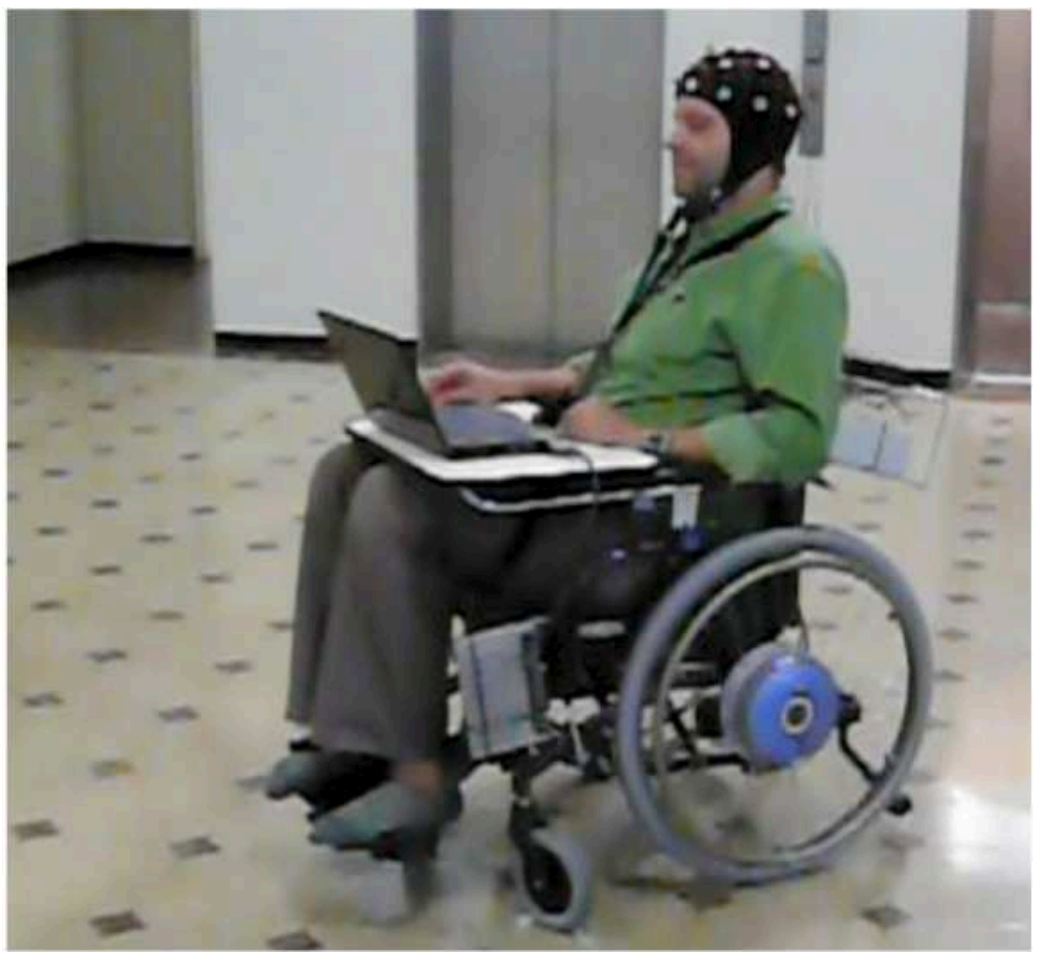

**Fig. 1. An example of classical visual BCI application for driving a wheelchair without any bodily movements. The subject has to fully concentrate on a display computer display without possibility to pay attention on the surroundings, which might cause a potential accident danger. The experiment was conducted in Laboratory for Advanced Brain Signal Processing, RIKEN Brain Science Institute, Japan.**

We propose to utilize the both spatial audio and tactile stimuli cues based designs with a target application in the new BCI paradigms where users intentionally direct their attention to different sounds and touch locations in multisensory stimulation environment. In order to identify user's target responses to presented auditory or tactile stimuli we first have to preprocess the EEG signals in order to decompose them into components carrying stimuli evoked potentials (so called event-related potentials). In order to achieve this we utilize a signal processing pipeline composed of EMD technique with spectral clustering followed by signals averaging and their locations estimation which, due to space limit constrains, is not discussed in this paper (refer to Rutkowski et al., 2010).

In the following sections we review the two auditory and tactile BCI techniques. The state of the art results summary and discussion conclude the paper.

## 2. SPATIAL AUDITORY BCI PARADIGMS

Spatial auditory BCI (saBCI) utilize human space cognition features as stimulus cues. This concept allows for natural and realistic virtual interaction environments design (Rutkowski et al., 2009; Schreuder et al., 2010; Rutkowski, 2011) in which user adopt easily using their natural auditory spatial perception. The auditory stimulus can be generated using the real sound sources (loudspeakers, etc.) or virtual auditory images (created with multimedia signal processing techniques allowing for positioning virtually sound sources arbitrarily between the real loudspeakers). In the following sections the both, real and virtual sound images based saBCI paradigms are described.

### 2.1 Real sound images based spatial stimulus saBCI

In the real sound images saBCI paradigm, the user interacts only with sound source stimuli that are generated using a sound environment consisting of multiple loudspeakers distributed spatially around a head. The spatially distributed loudspeakers are controlled by our in house developed multimedia environment in real time (Cai et al., 2012). Each loudspeaker can generate single real sound image stimulus separately, thus only sound originated from it (without any spatial virtualization) is considered (Cai et al., 2012; Cai et al., 2013). Figure 2 presents grand mean plots of brain responses captured during an eight commands saBCI interaction of nine users. The very clear "aha-responses," or so called P300 since they start 300 ms after each target stimulus onset and they have positive EEG deflection, are depicted there. It is thus possible to successfully implement the saBCI based on real sound images generated by the exactly same number of spatially distributed loudspeakers as the number of required BCI commands. A side effect of such design is an application complexity and a higher cost related to the larger number of the necessary loudspeakers. Thus, a possibility to generate virtual sound image stimuli with lower number of necessary loudspeakers is discussed in the next section.

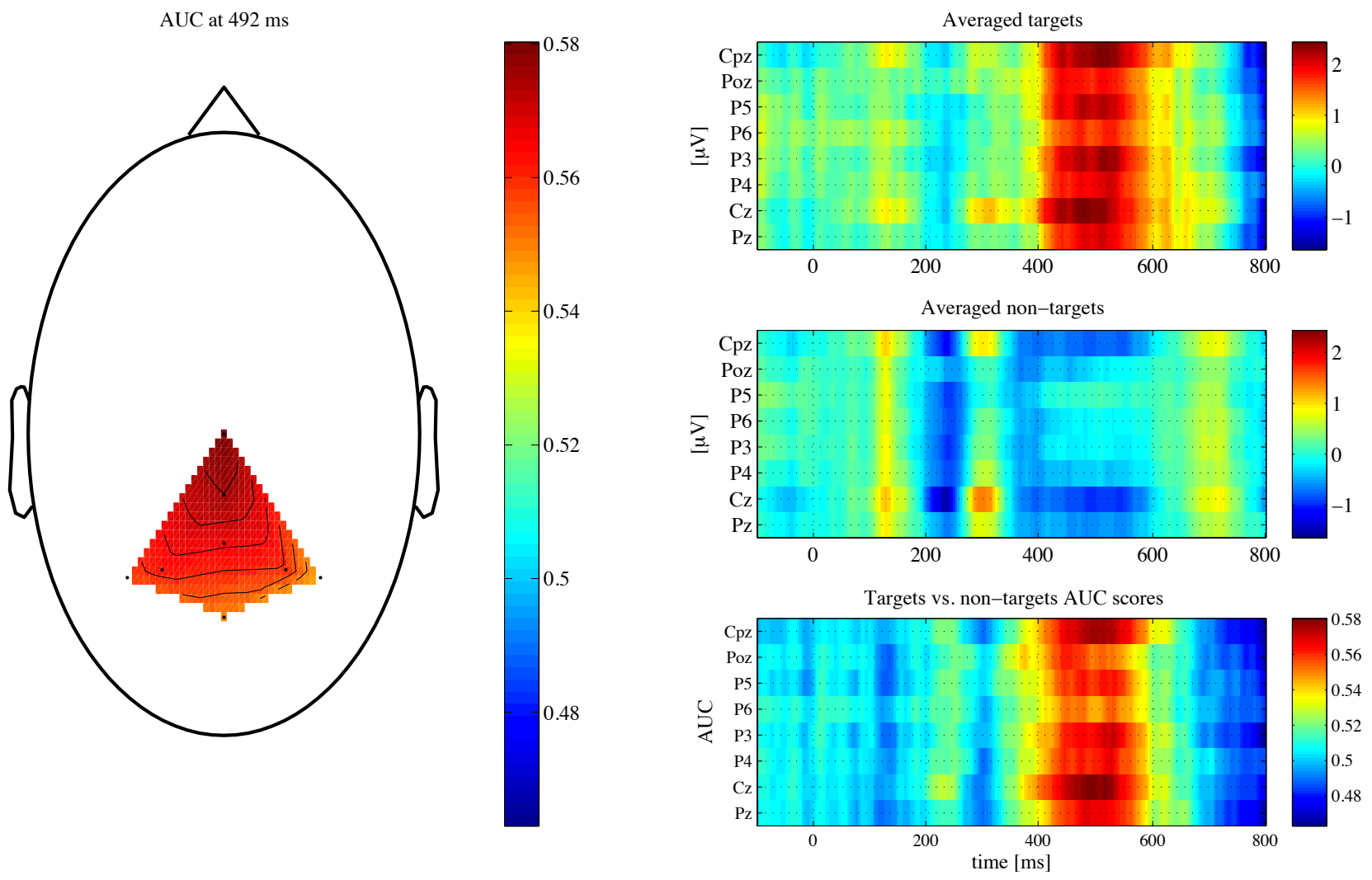

**Fig. 2. Grand mean EEG brain evoked potentials from the nine subjects in response to the real sound image stimuli. The right head topographical plot depicts the scalp locations to which the EEG electrodes were attached. All of the chosen electrodes responded with equal discriminative features at the maximum point estimated from area under the curve (AUC) feature depicted in right lower graph (Theodoridis and Koutroumbas, 2009). The right panel two upper plots depict evoked responses from the all EEG channels to attended targets (top) and non-targets (middle). The clear P300 or "aha-response" can be seen for targets in the range of 400-600 ms. AUC time series for all electrodes used are also depicted in the right lower bottom panel confirming the finding in grand mean responses visualized above.**

### 2.2 Virtual sound image localization based spatial stimulus saBCI

In the virtual sound images saBCI paradigm, the user receives virtualized sound source stimuli, which are created using a vector based amplitude panning (VBAP) approach (Pulkki, 1997; Nishikawa et al., 2012a; Nishikawa et al., 2012b). The VBAP application can be used to position virtually the sound image between the pairs of neighboring loudspeakers. For example, positioning a virtual sound image at a direction of 45° is possible with VBAP using the two loudspeakers placed at 0° and 90° that play the same stimulus with equal loudness simultaneously. In the utilized by the authors horizontal octagonal loudspeakers setup (Nishikawa et al., 2012a; Nishikawa et al., 2012b) it is possible to virtually position a sound image at any randomly chosen spatial location, which is the major strength of the approach allowing limitation of the necessary loudspeaker units. Figure 3 presents grand mean plots of brain responses captured during an eight commands saBCI interaction of nine users using the virtual sound images. Here again (compare with results in Figure 2) the very clear "aha-responses" or P300 responses are depicted. Therefore, it is also possible to successfully implement the saBCI based on virtual sound images generated by the lower number of spatially distributed loudspeakers comparing to the traditional real sound sources case.

## 3. SPATIAL TACTILE BCI PARADIGMS

Recently alternative to visual and auditory BCI paradigms solutions have been proposed to make use the brain somatosensory (tactile) channel to allow targeting of the tactile sensory domain for the operation of robotic equipment such as personal vehicles (Mori et al., 2013a), life support systems, etc. The rationale behind the use of the tactile channel is that it is normally far less loaded than visual or even auditory channels in such applications. A novel tactile brain-computer interface based on P300 responses evoked by tactile stimuli delivered via vibrotactile exciters attached to the ten fingertips of the subject's hands has been reported in (Mori et al., 2013a) and here we only review those results presented in Figure 4. This figure presents grand mean plots of brain responses captured during a ten commands saBCI interaction of seven users using the finger tactile stimuli. The very clear "aha-" or P300-responses are depicted there. Thus, this proves also that it is also possible to successfully implement the tactile BCI based on vibrotactile finger stimulation.

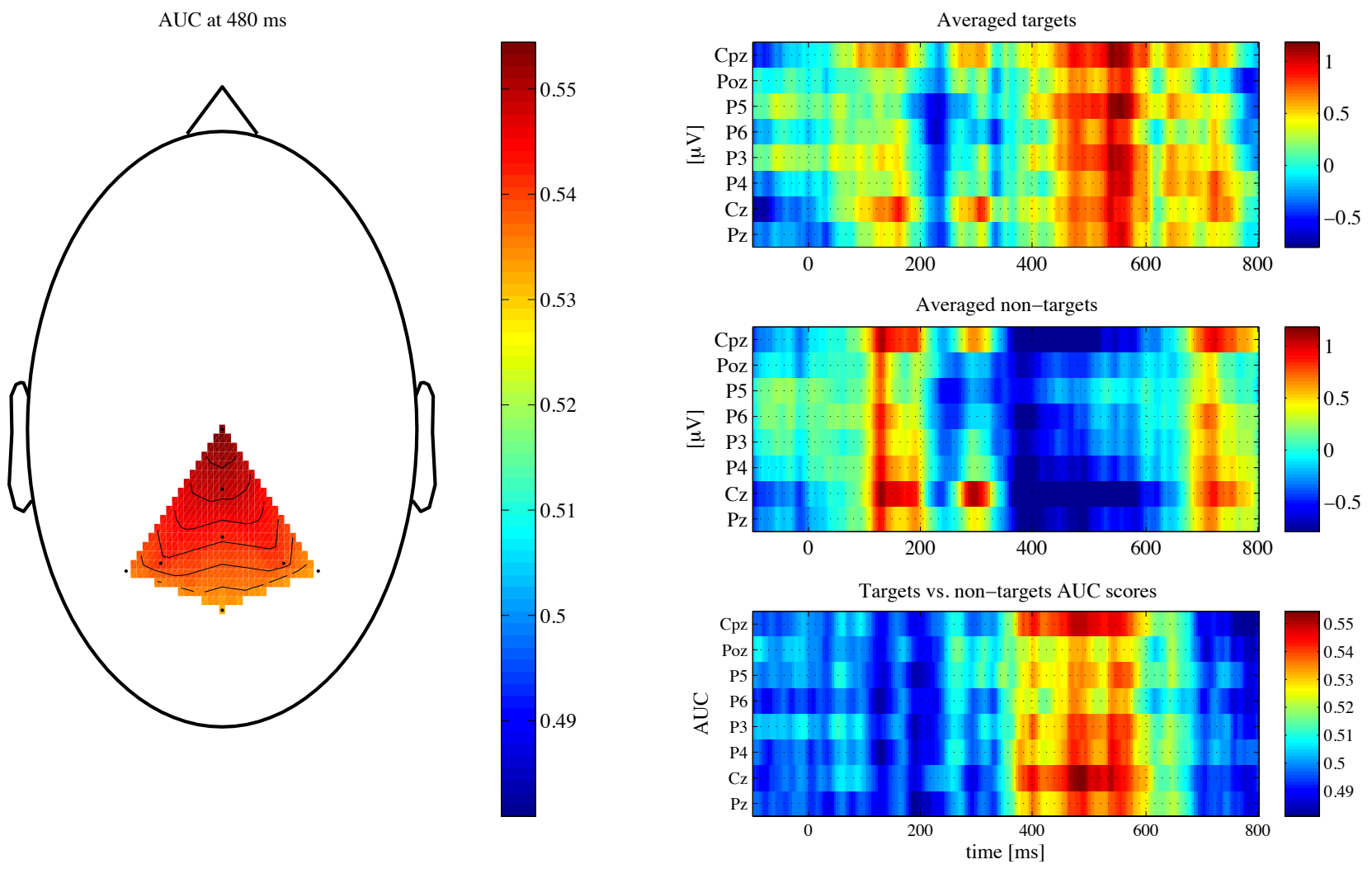

**Fig. 3. Grand mean averaged EEG brain evoked potentials from the nine subjects in response to the virtual sound image stimuli. The right head topographical plot depicts the scalp locations to which the EEG electrodes were attached. All of the chosen electrodes responded with equal discriminative features at the maximum point estimated from area under the curve (AUC) feature depicted in right lower graph (Theodoridis and Koutroumbas, 2009). The right panel two upper plots depict evoked responses from the all EEG channels to attended targets (top) and non-targets (middle). The weaker, comparing to result from Fig. 2, P300 or "aha" responses can be also seen for targets in the range of 400-600 ms. AUC time series for all electrodes used are also depicted in the right lower bottom panel confirming the finding in grand mean responses visualized above.**

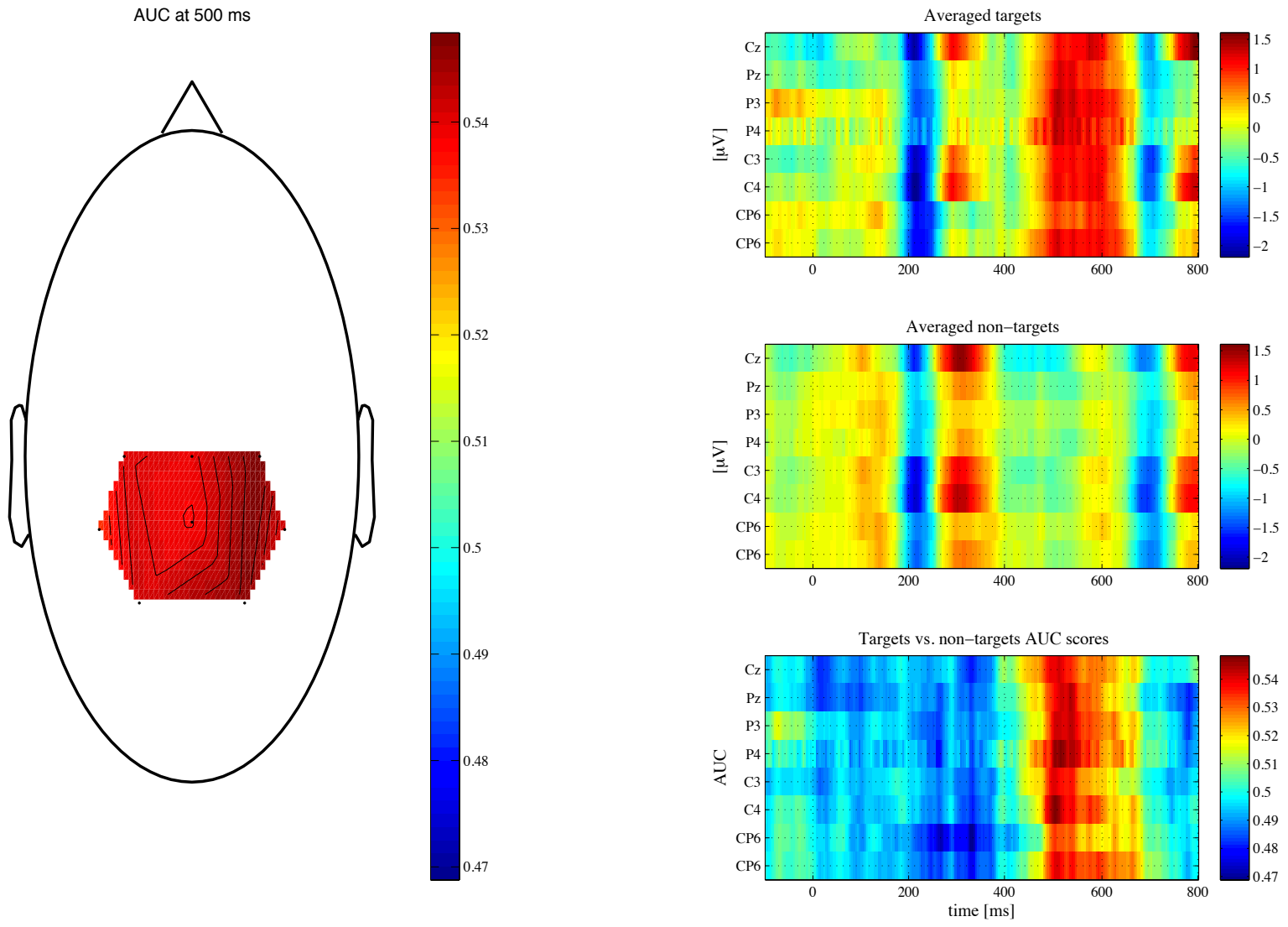

**Fig. 4. Grand mean averaged EEG brain evoked potentials from the seven subjects in response to the finger based tactile stimuli. The right head topographical plot depicts the scalp locations to which the EEG electrodes were attached. All of the chosen electrodes responded with equal discriminative features at the maximum point estimated from area under the curve (AUC) feature depicted in right lower graph (Theodoridis and Koutroumbas, 2009). The right panel two upper plots depict evoked responses from the all EEG channels to attended targets (top) and non-targets (middle). The P300 or "aha" responses can be also seen for targets in the range of 500-700 ms. AUC time series for all electrodes used are also depicted in the right lower bottom panel confirming the finding in grand mean responses visualized above.**

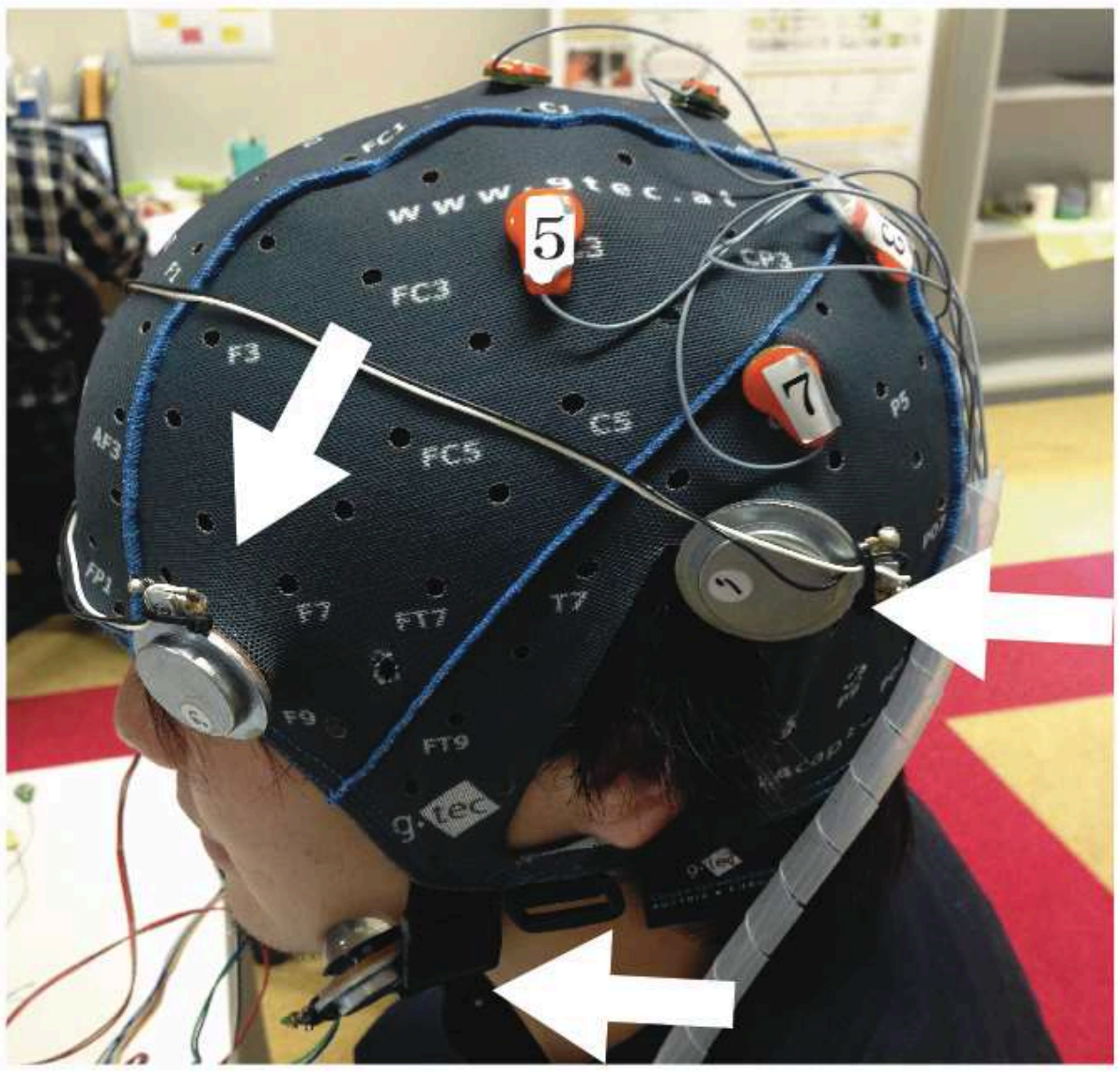

**Fig. 5. Example of vibrotactile and bone-conduction stimulators setup on a user head (marked with white arrows) together with dry EEG electrodes (orange dots with numbers)**

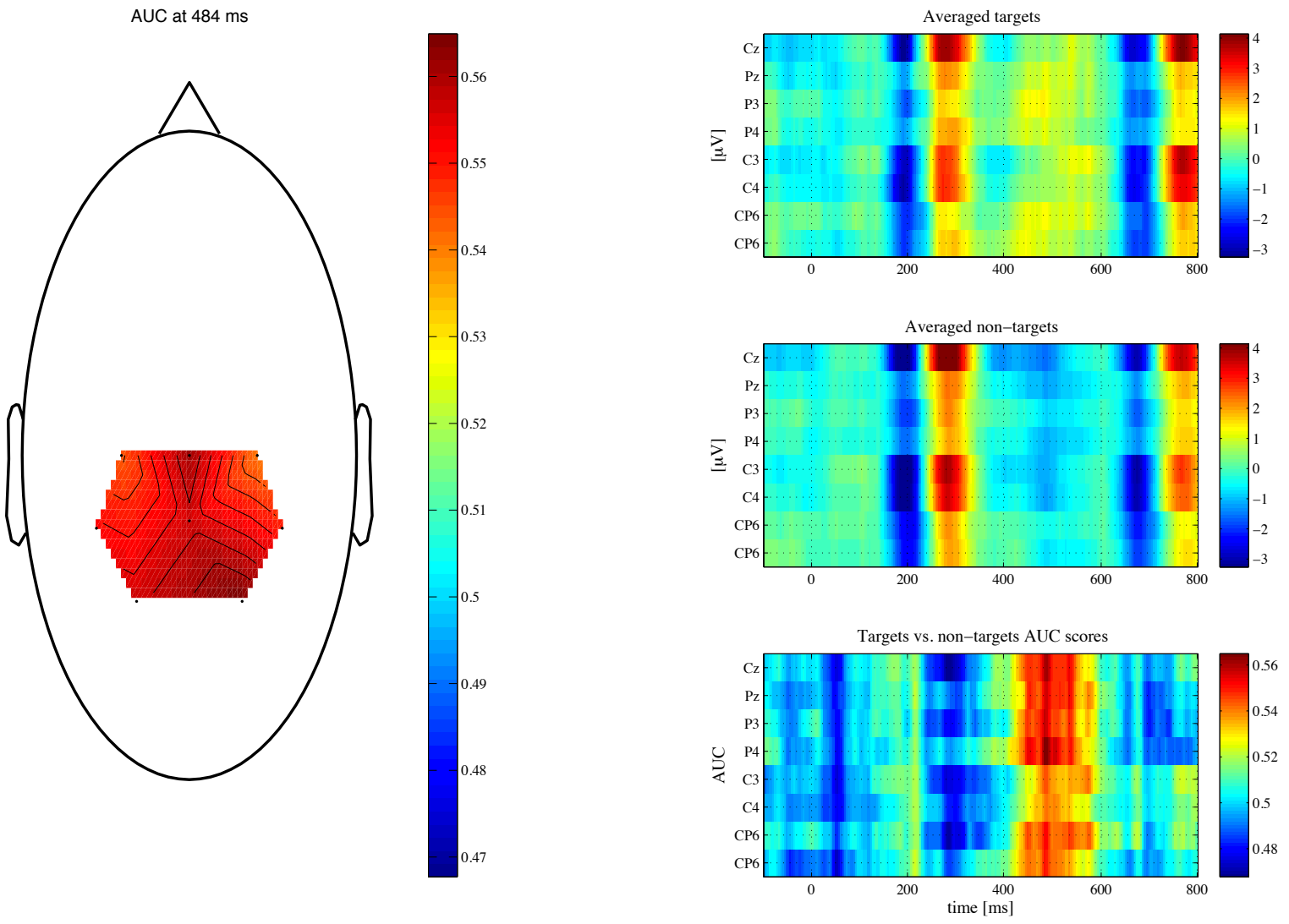

**Fig. 5. Grand mean averaged EEG brain evoked potentials from the seven subjects in response the head tactile and auditory (bone-conduction) based stimuli. The right head topographical plot depicts the scalp locations to which the EEG electrodes were attached. All of the chosen electrodes responded with equal discriminative features at the maximum point estimated from area under the curve (AUC) feature depicted in right lower graph (Theodoridis and Koutroumbas, 2009). The right panel two upper plots depict evoked responses from the all EEG channels to attended targets (top) and non-targets (middle). The P300 or "aha" responses can be also seen for targets in the range of 500-700 ms. AUC time series for all electrodes used are also depicted in the right lower bottom panel confirming the finding in grand mean responses visualized above.**

## 4. SPATIAL BIMODAL TACILE AND AUDITORY BCI PARADIGMS

The final BCI paradigms extension beyond classical visual paradigms is related vibrotactile stimuli delivered to the head of a subject which can serve as a novel interfacing platform. Six head positions are used to evoke combined somatosensory and auditory (via the bone conduction effect) brain responses, in order to define a multimodal tactile and auditory brain computer interface (taBCI). Experimental results of subjects performing online taBCI, using stimuli with a moderately fast inter-stimulus interval, has validate the taBCI paradigm, while the feasibility of the concept has been already illuminated through information transfer rate case studies (Mori et al., 2013b). An example of taBCI vibrotactile exciters and EEG dry electrodes setup is presented in Figure 6. Six vibrotactile exciters and eight electrodes are all attached to an EEG cap which the user comfortably wears on his/her head. Results of taBCI interaction of seven users in form of grand means are presented in Figure 7 confirming the paradigm usability in online application for healthy and paralyzed subjects.

## 5. CONCLUSIONS

If the results of the presented studies will not directly lead very soon to the successful commercial applications of spatial auditory and tactile allowing for an improvement and restoration of communication ability of totally-locked-in patients, they demonstrate that the use the new sensory modalities amend the interfacing choices based on the very good brain response results reported in this paper.

In the reviewed experimental results only the BCI-naïve subjects too part which confirms usability of the proposed solutions. The preliminary, yet encouraging results presented are a step forward in the search for new BCI paradigms for the paralyzed and totally-locked-in patients with compromised vision and hearing symptoms.

## 6. ACKNOWLEDGEMENT

This research was supported in part by the Strategic In- formation and Communications R&D Promotion Program no. 121803027 of The Ministry of Internal Affairs and Communication in Japan, and by KAKENHI, the Japan Society for the Promotion of Science, grant no. 12010738. We also acknowledge the technical support of YAMAHA Sound & IT Development Division in Hamamatsu, Japan.

## REFERENCES

[Brouwer and Van Erp, 2010] Brouwer, A.-M. and Van Erp, J. B. F. (2010). A tactile P300 brain-computer interface. Frontiers in Neuroscience, 4(19).

[Cai et al., 2012] Cai, Z., Makino, S., and Rutkowski, T. M. (2012). Spatial auditory BCI paradigm utilizing N200 and P300 responses. In Proceedings of the Fourth APSIPA Annual Summit and Conference (APSIPA ASC 2012), page paper #355, Hollywood, CA, USA. APSIPA.

[Cai et al., 2013] Cai, Z., Makino, S., and Rutkowski, T. M. (2013). Brain evoked potential latencies optimization for spatial auditory brain-computer interface. Cognitive Computation, pages (accepted, in press).

[Chang et al., 2013] Chang, M., Nishikawa, N., Struzik, Z. R., Mori, K., Makino, S., Mandic, D., and Rutkowski, T. M. (2013). Comparison of P300 responses in auditory, visual and audiovisual spatial speller BCI paradigms. In Proceedings of the Fifth International Brain-Computer Interface Meeting 2013, page Article ID: 156, Asilomar Conference Center, Pacific Grove, CA USA. Graz University of Technology Publishing House, Austria.

[Mori et al., 2012] Mori, H., Matsumoto, Y., Makino, S., Kryssanov, V., and Rutkowski, T. M. (2012). Vibrotactile stimulus frequency optimization for the haptic BCI prototype. In Proceedings of The 6th International Conference on Soft Computing and Intelligent Systems, and The 13th International Symposium on Advanced Intelligent Systems, pages 2150-2153, Kobe, Japan.

[Mori et al., 2013a] Mori, H., Matsumoto, Y., Kryssanov, V., Cooper, E., Ogawa, H., Makino, S., Struzik, Z., and Rutkowski, T. M. (2013a). Multi-command tactile brain computer interface: A feasibility study. In Oakley, I. and Brewster, S., editors, Haptic and Audio Interaction Design 2013 (HAID 2013), volume 7989 of Lecture Notes in Computer Science, pages 50–59. Springer Verlag Berlin Heidelberg.

[Mori et al., 2013b] Mori, H., Matsumoto, Y., Struzik, Z. R., Mori, K., Makino, S., Mandic, D., and Rutkowski, T. M. (2013b). Multi-command tactile and auditory brain computer interface based on head position stimulation. In Proceedings of the Fifth International Brain-Computer Interface Meeting 2013, page Article

ID: 095, Asilomar Conference Center, Pacific Grove, CA USA. Graz University of Technology Publishing House, Austria.

[Krusienski et al., 2006] Krusienski, D. J., Sellers, E. W., Cabestaing, F., Bayoudh, S., McFarland, D. J., Vaughan, T. M., and Wolpaw, J. R. (2006). A comparison of classification techniques for the P300 speller. Journal of Neural Engineering, 3(4):299.

[Ortner et al., 2012] Ortner, R., Rodriguez, J., and Guger, C. (2012). A tactile P300 brain-computer inter- face. Technical Report 3, g.tec Medical Engineering GmbH, Austria.

[Nishikawa et al., 2012a] Nishikawa, N., Makino, S., and Rutkowski, T. M. (2012a). Analysis of brain responses to spatial real and virtual sounds - a BCI/BMI approach. In Bandyopadhyay, A., editor, Proceedings of The Second International Workshop on Brain Inspired Computing (BIC2012), page poster #2.3 (The Best Poster Award), Tsukuba, Japan. NIMS.

[Nishikawa et al., 2012b] Nishikawa, N., Matsumoto, Y., Makino, S., and Rutkowski, T. M. (2012b). The spatial real and virtual sound stimuli optimization for the auditory BCI. In Proceedings of the Fourth APSIPA Annual Summit and Conference (APSIPA ASC 2012), page paper #356, Hollywood, CA, USA. APSIPA.

[Pulkki, 1997] Pulkki, V. (1997). Virtual sound source positioning using vector base amplitude panning. Journal of Audio Engineering Society, 45(6):456–466.

[Rutkowski et al., 2009] Rutkowski, T. M., Cichocki, A., and Mandic, D. P. (2009). Spatial auditory paradigms for brain computer/machine interfacing. In International Workshop On The Principles and Ap- plications of Spatial Hearing 2009 (IWPASH 2009) - Proceedings of the International Workshop, page P5, Miyagi-Zao Royal Hotel, Sendai, Japan.

[Rutkowski et al., 2010] Rutkowski, T. M., Mandic, D. P., Cichocki, A., and Przybyszewski, A. W. (2010). EMD approach to multichannel EEG data - the amplitude and phase components clustering analysis. Journal of Circuits, Systems, and Computers (JCSC), 19(1):215–229.

[Rutkowski, 2011] Rutkowski, T. M. (2011). Auditory brain-computer/machine interface paradigms design. In Cooper, E., Kryssanov, V., Ogawa, H., and Brewster, S., editors, Haptic and Audio Interaction Design, volume 6851 of Lecture Notes in Computer Science, pages 110–119. Springer Berlin / Heidelberg.

[Rutkowski et al., 2013] Rutkowski, T. M., Mori, H., Matsumoto, Y., Struzik, Z. R., Makino, S., Mandic, D., and Mori, K. (2013). Spatial tactile and auditory brain computer interface based on head position stimulation. In Abstracts of the 36th Annual Meeting of the Japan Neuroscience Society (Neuro2013), pages O2–6–3–3, Kyoto, Japan. Japan Neuroscience Society.

[Schreuder et al., 2010] Schreuder, M., Blankertz, B., and Tangermann, M. (2010). A new auditory multi- class brain-computer interface paradigm: Spatial hearing as an informative cue. PLoS ONE, 5(4):e9813.

[Theodoridis and Koutroumbas, 2009] Theodoridis, S. and Koutroumbas, K. (2009). Pattern Recognition. Acedemic Press, fourh edition.

[van der Waal et al., 2012] van der Waal, M., Severens, M., Geuze, J., and Desain, P. (2012). Introducing the tactile speller: an ERP-based brain–computer interface for communication. Journal of Neural Engineering, 9(4):045002.

[Wolpaw and Wolpaw, 2012] Wolpaw, J. and Wolpaw, E. W., editors (2012). Brain-Computer Interfaces: Principles and Practice. Oxford University Press.